\documentclass[10pt]{article}
\input epsf
\begin{document}

\title{Relativistic non-equilibrium thermodynamics revisited}
\author{L.S. Garc\'{\i}a-Col\'{\i}n$^{a,\,b}$ and A. Sandoval-Villalbazo$^c$\\
$^a$ Departamento de F\'{\i}sica, Universidad Aut\'{o}noma Metropolitana \\
M\'{e}xico D.F., 09340 M\'{e}xico \\
$^b$ El Colegio Nacional, Centro Hist\'{o}rico 06020 \\
M\'{e}xico D.F., M\'{e}xico \\
E-Mail: lgcs@xanum.uam.mx \\
$^c$ Departamento de F\'{\i}sica y Matem\'{a}ticas, Universidad Iberoamericana \\
Lomas de Santa Fe 01210 M\'{e}xico D.F., M\'{e}xico \\
E-Mail: alfredo.sandoval@uia.mx \\} \maketitle
\bigskip

\begin{abstract}
Relativistic irreversible thermodynamics is reformulated following
the conventional approach proposed by Meixner in the
non-relativistic case. Clear separation between mechanical and
non-mechanical energy fluxes is made. The resulting equations for
the entropy production and the local internal energy have the same
structure as the non-relativistic ones. Assuming linear
constitutive laws, it is shown that consistency is obtained both
with the laws of thermodynamics and causality.
\end{abstract}

\bigskip

\section{Introduction}
Classical or non-equilibrium thermodynamics (LNT) as formulated
originally by J. Meixner \cite{Meixner1} \cite{Groot1}  and given
its final and more accessible version in the classical monograph
by S.R. de Groot and P. Mazur \cite{kin1} first published in 1961
is, at present, the most complete and self consistent framework
available to deal with non-equilibrium phenomena. It is firmly
rooted in both statistical mechanics \cite{Steele}  and in the
kinetic theory of gases based on the Boltzmann equation
\cite{Groot1} \cite{kin2} \cite{GC1}. It is also well known that
the theory has its limitations, specially when the so-called
constitutive equations fail to describe correctly phenomena which
occur in the presence of large gradients or when memory effects
are non-negligible.The latter issue is well known from the work of
Kohlrausch and Weber in the first half of the nineteenth century,
dealing with responses of systems like glass fibers and similar
systems \cite{GC2}. These observations are readily available today
in rheological polymeric fluids,all types of glasses forming
liquids in the neighborhood of the glass transition temperature,
and others. The extension of LNT to deal with such phenomena has
given rise to what is generically known as Extended
Non-equilibrium Thermodynamics (ENT). Of about the so far seven
different versions of this theory \cite{Jou1} \cite{GC3}
\cite{Jou2} non of them is completely convincing.

The main object of this paper is addressed to a natural question
namely the relativistic version of LNT. Some authors have jumped
to conclusions as to why something like the relativistic version
of one of the existing theories belonging to ENT has to be used,
specially in some astrophysical and cosmological phenomena without
really understanding clearly the tenets of the former one.
Although the reason for this attitude is easy to understand, what
is astonishing is that the relativistic version of Meixner's
original theory has been only attempted in one occasion before
\cite{Nos1} and, unfortunately, in a very clumsy representation.
This is why in this paper we wish to go back to this question with
the idea of convincing the reader that relativistic LNT is far
more powerful than what has been hinted in previous works and that
it may be regarded as consistent, both with the tenets of the
theory of relativity and the first two laws of thermodynamics.

To clarify ideas lets go back to their origin. In 1940 C. Eckart
published three papers entitled \emph{The thermodynamics of
irreversible processes} \cite{Eckart}, the first two dealing, in
much the same fashion done by Meixner, with a simple fluid, fluid
mixtures and the third one addressing the problem of a
relativistic simple fluid. This happened just a few years before
Meixner published his own full version of LNT which was first
communicated in English by de Groot in 1952 \cite{Groot1} and
finally brought to its present version twenty years later by de
Groot and Mazur \cite{kin1}. The point is that in his third paper,
Eckart proceeded in the standard way except for the fact that he
faced the problem of relating the first law of thermodynamics with
the standard energy-momentum tensor of relativity, and, in his own
words, could not find a heat density to combine it with the three
vector associated with the flow of heat to form a four-vector.
This is precisely the root of one of the difficulties found today
with the papers dealing with this subject. In one way or another,
several authors have followed Eckart,  who proposed to construct a
tensor which would include both the internal energy and the flow
of heat. As  had been pointed  out already by other authors, this
is against the tenets of the general theory of relativity, the
stress energy tensor includes only all forms of mechanical energy
\cite{Tolman} \cite{Weyl}. Heat cannot be incorporated into its
structure. It is surprising that most of the papers written today
on this questions keep at all cost this point of view. Since
Eckart's theory leads to results that violate causality and
involves undesirable unstable modes, it has been patched up in
several ways using the ideas introduced by Israel and coworkers
\cite{Israel1} \cite{Israel2} \cite{Israel3} and even resorting to
some of the versions of EIT \cite{Zimdahl} \cite{Jou4}.

There is one last issue  of upmost importance to deal with, what
authors in this field refer to as the "order" in the theories. The
most common statement is to disqualify a theory if the entropy
current contains only terms of first order in the deviations from
equilibrium \cite{Zimdahl} \cite{Jou4} \cite{Hiscock}. This is
kind of meaningless. First of all one has to remember that the
entropy balance equation is solely and uniquely a consequence of
the local equilibrium hypothesis and the balance equations. The
former one is by no means whatsoever a well established criteria
to determine wether or not a system exhibits behavior which is
close or far from equilibrium. Needless to mention here examples,
but just to clarify matters we can recall the Burnett equations of
hydrodynamics which as shown recently \cite{GC4} \cite{GC5} are
the best option describing the structure of shock waves at high
Mach numbers and are perfectly consistent with the local
equilibrium assumption. It is the generalization of the linear (in
the gradients) constitutive equations which account for this
improvement. The entropy current in completely invariant to this
feature and is still given its standard form:
\begin{equation}
\vec{J_{S}}=\frac{\vec{J_{Q}}}{T}+\rho s \vec{u}
\end{equation}
where $\vec{J_{Q}}$ is the heat flow vector, $T$ and $\vec{u}$ the
local values of the temperature and velocity, respectively, and
$\rho s$ the local entropy per unit of volume. Clearly, since now
$\vec{J_{Q}}$ obeys a constitutive equation much more complicated
than that arising from Fourier's equation,  the explicit form for
$\vec{J_{S}}$ is such that it contains terms of order higher than
the first in the gradients, and issue ignored by most authors.
Thus, the question of "high order deviations from equilibrium"  is
completely foreign to the generic form of $\vec{J_{S}}$.

All the features pointed out above are the topic of this paper.
Indeed, if one follows the canonical rules behind the standard
theory of LNT using the basic principles of the theory of
relativity one obtains a set of equations of motion for the chosen
local state variables whose order on the gradients arise because
of the constitutive equations, which we underline, are foreign to
the theory and that one chooses to relate the fluxes with the
forces. The linear laws give rise the Navier-Stokes-Fourier
equations of hydrodynamics in the case of a simple fluid which, by
the way, are already non-linear in the gradients. Higher order in
the gradients may be studied by using what we could call the
general Burnett constitutive equations leading to what some
authors refer to as "second order in the deviations from
equilibrium", but a full discussion of these equations will be
undertaken later. Lastly, we also bring to the fore another
ignored feature inherent to a relativistic thermodynamic theory,
namely, necessary and sufficient conditions that the linear
constitutive equations must fulfill so that the theory obeys
causality and consistency with the second law of thermodynamics.

To accomplish our task we have structured the paper as follows. In
section two the basic assumptions behind the relativistic version
of classical non-equilibrium thermodynamics is discussed in a very
simple representation. In section 3 we derive the hydrodynamic or
transport heat equation for a simple shear free fluid. Shearing
stresses may be trivially included but we ignore them for
pedagogical reasons. It is at this stage where consistency with
causality and the second law enter into the formalism. A
discussion on the results and some concluding remarks are left to
section 4.

\section{Relativistic linear non-equilibrium thermodynamics}

As we mentioned in the introduction, our first task in this paper
is to use the structure of the conservation equations together
with the local equilibrium assumption to derive an equation for
the entropy balance using a representation in which the
independent local variables are taken to be the number of
particles per unit of volume $n(x^{i},t)$, the four velocity
$u^{\nu}$ ($\nu=1,...,4$) and the internal energy per particle
$\varepsilon(x^{i},t)$. The calculation will be done within the
framework of the general theory of relativity assuming that the
fluid is isotropic and free from shearing stresses, that is shear
viscosity is neglected. Therefore, the mass-stress tensor has the
form:
\begin{equation}
\label{cero} T^{\mu}_{\nu}=\rho u^{\mu} u_{\nu}+\tilde{p}
h^{\mu}_{\nu}
\end{equation}
where $\rho$ is the mass density and $\tilde{p}=p+ p' $,  $p$
being the local hydrostatic pressure and $p'$ the diagonal part of
the stress tensor, namely the one responsible for bulk stresses.
Also, $h^{\mu}_{\nu}$ by many authors referred to as the spatial
projector for reasons to be discussed later, is defined as
\begin{equation}
\label{uno}
h^{\mu}_{\nu}=\delta^{\mu}_{\nu}-\frac{1}{c^{2}}u^{\mu} u_{\nu}
\end{equation}
so that $u^{\mu} u_{\mu}=c^{2} $,  $u_{\mu} h^{\mu}_{\nu}=0$ and $
u^{\nu} h^{\mu}_{\nu ;\mu}=-\theta$, where $\theta \equiv
u^{\mu}_{;\mu}$.

Since the basic conservation equation reads as
\begin{equation}
\label{dos} T^{\mu}_{\nu ;\mu}=0
\end{equation}
straightforward algebra leads to the result that
\begin{equation} \label{tres}
(\rho-\frac{\tilde{p}}{c^{2}})\dot{u}_{\nu}+u_{\nu}(\dot{\rho}+\rho
\theta-\frac{p}{c^{2}}\theta)=-\tilde{p}_{,\mu}h^{\mu}_{\nu}
\end{equation}
where $\dot{u_{\nu}}=u^{\mu} u_{\nu ;\mu}$. Multiplication of both
sides of Eq. (\ref{tres}) with $ u^{\nu} $ yields the mechanical
energy balance equation namely,
\begin{equation} \label{cuatro}
\dot{\rho} c^{2}+\rho c^{2} \theta -\tilde{p} \theta=0
\end{equation}
On the other hand, if we assume that the number of particles is
conserved and defining the particle flux as
\begin{equation} \label{cinco}
N^{\mu}=n u^{\mu}
\end{equation}
such conservation requirement is met by the condition that
\begin{equation} \label{seis}
N^{\mu}_{;\mu}=\dot{n}+n \theta=0
\end{equation}
Eqs. (\ref{cuatro}) and \ref{seis}) will be particularly useful in
the construction of balance equations for internal energy per
particle $\varepsilon$ as well as for the entropy per unit of
volume $ns$. To do so we recall that in Meixner's formulation of
classical irreversible thermodynamics
\cite{Meixner1}-\cite{Groot1} one assumes that the local total
energy density is conserved. This implies that the total energy
flux contains both the flux of mechanical energy plus the
non-mechanical one, namely the heat flux. This statement is
equivalent to assuming the validity of the first law of
thermodynamics \cite{kin1}. Thus, we write such flux as:

\begin{equation} \label{siete}
J^{\mu}_{[T]}=u^{\nu} T^{\mu}_{\nu}+ n \varepsilon
u^{\mu}+J^{\mu}_{[Q]}
\end{equation}
where clearly,  $u^{\nu} T^{\mu}_{\nu}=\rho u^{\mu} c^{2}$ is the
\textit{mechanical energy flux}, $n \varepsilon u^{\mu}$ is the
internal energy flux and $J^{\mu}_{[Q]}$, a four vector, is the
heat flux.

Total energy conservation now requires that $J^{\mu}_{[T]
;\mu}=0$, whence from Eqs. (\ref{cuatro}), (\ref{seis}) and
(\ref{siete}), it follows that

\begin{equation} \label{ocho}
n \dot{\varepsilon}= -\tilde{p} \theta -J^{\mu}_{[Q] ;\mu}
\end{equation}
is the sought result for the internal energy balance equation. Eqs
(\ref{siete}) and (\ref{ocho}) require some additional remarks
since they are not always acknowledged nor understood.

The controversy regarding the correct method to deal the
relativistic properties of a simple fluid  originated from the
pioneering paper on the subject written by C. Eckart in 1940
mentioned before \cite{Eckart}. The main issue is that in this
paper the stress-energy tensor was constructed allowing the
inclusion of heat flow and internal energy. This seems to be
against the tenets of general relativity, as emphasized in the
introduction. The energy-momentum tensor is the most general
expression involving mechanical energy and, as emphasized in
Tolman's book \cite{Tolman}, contains no room for the first law of
thermodynamics. In fact, Eckart's version of Eq. (\ref{ocho})
contains a term of the form $\frac{1}{c} q^{\alpha}
\dot{u_{\alpha}}$, $q^{\alpha}$ representing the heat flow. This
term, difficult to interpret, is identified with a heat flow of
accelerated matter and is foreign to the structure of classical
irreversible thermodynamics as we shall see below.

Continuing with our argument, we now derive the entropy balance
equation using the local equilibrium assumption \cite{Groot1}
namely, the entropy per particle is a time independent functional
of $n$ and $\varepsilon$,
\begin{equation} \label{nueve}
s= s(n,\varepsilon)
\end{equation}
so that the Gibbs relation reads
\begin{equation} \label{diez}
n\dot{s}=\frac{n}{T}\dot{\varepsilon}-\frac{p}{n T}\dot{n}
\end{equation}
where the differential coefficients $(\frac{\partial s} {\partial
n})_{\varepsilon}$ and $(\frac{\partial s} {\partial
\varepsilon})_{n}$ have been evaluated using once more the local
equilibrium assumption. Direct substitution of Eqs. (\ref{ocho})
and (\ref{diez}) into Eq. (\ref{once}) yields immediately that

\begin{equation} \label{once}
n\dot{s}+(\frac{J^{\mu}_{[Q]}}{T})_{;\mu}=-\frac{J^{\mu}_{[Q]}T_{,
\mu}}{T^{2}}-\frac{p' \theta}{T}
\end{equation}
after a slight rearrangement of the resulting equation. Here $T$
is the local equilibrium temperature. The reader has to recognize
that this is a very encouraging result. The entropy balance
equation in general relativity using the $n$, $\varepsilon$
representation is identical in structure to the canonical
classical equation. The entropy flow is simply the divergence of
the ratio $\frac{J^{\mu}_{q}}{T}$ and the entropy production, here
represented by $\sigma$ is given by

\begin{equation} \label{doce}
\sigma=-\frac{J^{\mu}_{[Q]}T_{, \mu}}{T^{2}}-\frac{p' \theta}{T}
\end{equation}

Eq. (\ref{doce}) clearly expresses the full meaning of Clausius
uncompensated heat, it is the sum of the products between the
"macroscopic" gradients of the intensive variables $T$ and
$u^{\nu}$ and the corresponding flows they generate. This result,
together with Eq. (\ref{ocho}) are the main results of the first
part of this paper. Once more, these results differ from those
obtained by Eckart not only in the methodology used by this
author, but by the appearance of a term $\frac{1}{c T} q^{\alpha}
\dot{u{\alpha}}$, which may hardly be interpreted as a product of
a thermodynamic force and its corresponding flow. It is also worth
noting that Eckart's formalism cannot be naturally reconciled with
the Onsager's symmetry relations, while the canonical classical
equation shows this desirable feature.

What we have therefore accomplished in this section may be safely
considered as being the natural extension of classical
irreversible thermodynamics to the context of the general theory
of relativity. In many ways the results are an improvement over
these obtained by the same authors about ten years ago \cite{Nos1}
using the mass density $\rho$ as an independent variable instead
of $n$. The interested reader may look at the analogous of Eqs.
(\ref{diez}) and (\ref{doce}) as written in those papers.

We now go on to the next important question raised in the
introduction namely the structure and meaning of the transport
equations which arise from these theory. To fix our attention we
shall concentrate on the heat transport equation leaving the
momentum transport equation for future work.

\section{The heat transport equation}
Transport equations arise when the unknown quantities namely, the
fluxes appearing in the conservation equations are expressed in
terms of the independent variables $n$, $u^{\alpha}$ and
$\varepsilon$ in our case. This is achieved through the so-called
constitutive equations which, as is well known, are foreign to the
theory. They must be extracted either from experiment or from a
microscopic model, this latter possibility being rather difficult
to achieve except for some very simple systems. In our equations,
the unknowns are $p'$, the bulk momentum flow , and $J^{\mu}_{q}$
the heat flux vector. Also, due to practical reasons, it is useful
to eliminate $\varepsilon$ in terms of the local temperature, a
much more accessible variable, and $n$. This is allowed by the
local equilibrium hypothesis since one may always write that
$\varepsilon=\varepsilon(n,T)$, so that
\begin{equation} \label{trece}
n \dot{\epsilon}=n(\frac{\partial \epsilon}{\partial n})_{T}
~\dot{n}+n(\frac{\partial \epsilon}{\partial T})_{n} ~\dot{T}
\end{equation}
the thermodynamical coefficients in Eq. (\ref{trece}) are given
by:
\begin{equation} \label{catorce}
(\frac{\partial \epsilon}{\partial n})_{T}=-\frac{T \beta}{n^{2}
\kappa_{T}}+\frac{p}{n^{2}}
\end{equation}
and
\begin{equation} \label{quince}
(\frac{\partial \epsilon}{\partial T})_{n}=C_{n}
\end{equation}

In Eqs. (\ref{catorce}) and (\ref{quince}) $\beta$ is the thermal
expansion coefficient, $\kappa_{T}$ is the isothermal
compressibility and $C_{n}$ is the specific heat at constant
numerical density.  Eq.(\ref{trece}) transforms into

\begin{equation} \label{dieciseis}
(\frac{\partial \epsilon}{\partial T})_{n}=C_{n}
\dot{T}-\frac{p}{n} \dot{n}
\end{equation}
which, when combined with Eqs. (\ref{seis}) (\ref{ocho})  leads to
an equation for $\dot{T}$

\begin{equation} \label{diecisiete}
 -\tilde{p} \theta -J^{\mu}_{[Q] ;\mu}=(\frac{T \beta}{
\kappa_{T}}+p)  \theta +  C_{n} \dot{T}
\end{equation}
where now $p'$ and $J^{\mu}_{q ;\mu}$ have to be expressed in
terms of $u^{\nu}$, $T$ and $n$ through constitutive equations.
This brings us to a rather delicate question in all this
formalism. In fact, if we were to proceed according to the
postulates of Meixner's theory there should be a coupling, linear,
between fluxes and forces. One way to write these relations is

\begin{equation} \label{dieciocho}
p'= - \eta_{B} \theta
\end{equation}
where $\eta_{B}$ is the bulk viscosity coefficient and
\begin{equation} \label{diecinueve}
J^{\mu}_{[Q]}=-K g^{\mu \alpha} T_{,\alpha}
\end{equation}
where $K$ is the heat conductivity.  The reader may wonder about
the time component of Eq. (\ref{diecinueve}), since the absence of
a projector operator implies a kind of energy density associated
with the heat flux. It is interesting to notice that, indeed, even
in non-relativistic irreversible thermodynamics a non-vanishing
density energy associated to heat is needed to recover the first
law of thermodynamics from the internal energy balance equation.
This issue is discussed in Ref. \cite{kin1}. Substitution of Eqs.
(\ref{dieciocho}) and (\ref{diecinueve}) into Eq.
(\ref{diecisiete}) leads to the equation
\begin{equation} \label{veinte}
 \eta_{B} \theta^{2} + (K g^{\mu \alpha} T_{,\alpha})_{; \mu}=(\frac{T \beta}{
\kappa_{T}}+p)  \theta +  n C_{n} \dot{T}
\end{equation}
Since $\theta^{2}$ is a quadratic form in $u_{\nu ;\nu}$ and
$\frac{T \beta}{\kappa_{T}}$ is usually a small number for
ordinary gases, neglecting these last two terms we get that

\begin{equation} \label{veintiuno}
(K g^{\mu \alpha} T_{,\alpha})_{; \mu}= n C_{n} \dot{T}
\end{equation}
which is a hyperbolic type equation for $T$. Causality is not
violated and no temperature perturbations can propagate with a
velocity larger than $c$. Moreover, these constitutive equations
lead to an entropy production

\begin{equation} \label{veintidos}
\sigma=\frac{k T^{; \mu} T_{; \mu}}{T^2}+\frac{\eta_{v}
\theta^2}{T}
\end{equation}
which is a non-negative quadratic form since the transport
coefficients are known to be positive. Therefore, $\sigma>0$, in
complete agreement with the second law of thermodynamics. There
is, however, one problem. For isotropic homogeneous systems often
encountered in cosmological applications, the spatial gradients
vanish and Eq. (\ref{diecinueve}) would imply that

\begin{equation} \label{veintitres}
J^{4}_{[Q]}=-\frac{K}{c^{2}} \frac{\partial T}{\partial t}
\end{equation}

This implies that the fourth component of the heat flow
four-vector seems to "dissipate in time". The coefficient
$\frac{K}{c^{2}}$ will in general be a very small number, but not
zero.

Many authors do not accept this argument and keep the constitutive
equations strictly spatially projected. For this purpose, it is
proposed that Eq. (\ref{diecinueve}) should read as:

\begin{equation} \label{veinticuatro}
J^{\mu}_{[Q]}=-K h^{\mu \alpha} T_{,\alpha}
\end{equation}
so, when $h^{^{4 \alpha}} T_{,\alpha}$ is computed in the
co-moving system it is trivial to see that its value is zero so
that indeed Eq. (\ref{veintidos}) reduces to the ordinary
Fourier's equation. However, the second order in time derivative
disappears and the counterpart of Eq. (\ref{veintiuno}) becomes
parabolic.

Thus, we have two alternatives. One is to accept some kind of
dissipation along the time axis in the four dimensional space.
This guarantees causality and consistency with the second law
\cite{Nos4}. The other one is to project out from the constitutive
equations the temporal components of the fluxes and reduce them to
their classical expressions, loosing causality end direct
consistency with the second law. These features of both
formulations are perhaps the main reason as to why relativistic
non-equilibrium thermodynamics has been somewhat ignored.

\section{Concluding remarks}

As we have clearly shown in the previous sections, a relativistic
generalization of classical non-equilibrium thermodynamics can be
achieved without using arguments which are foreign to the Meixner
scheme. The entropy balance equation arises solely form the
conservation equations and the local equilibrium assumption. The
introduction of linear constitutive equations yield a positive
entropy production in agreement with the second law.
Generalization of these constitutive equations, which are foreign
to the theory, may be done consistently with the local equilibrium
assumption, but the \textit{local} positive definiteness of
$\sigma$ is lost. This already occurs in the non-relativistic
limit, as has been extensively discussed in the literature that
the Burnett and higher order corrections do not yield a local
positive entropy production, but only a global one. This is the
main difference between our approach and the ones followed by
Israel and collaborators and more recently by Pavon, Zimdahl and
others \cite{Israel3}-\cite{Jou4}. They solve the problem using
ideas of ENT incorporating fluxes themselves as state variables.
Hence, the entropy of the system cannot be defined \cite{GC2}
\cite{GC3} as well as the second law. So, the quadratic  terms in
the entropy flux proposed one way or the other  by these authors,
postulating Maxwell-Cattanneo type equations as constitutive
relations leads to a theory in which neither Clausius entropy nor
the second law in its conventional form have a place on their own.

The other remark comes from the results obtained some time ago by
Hiscock and Lindblom  \cite{Hiscock} and fully availed by Israel
\cite{Israel4} asserting that the relativistic version of the
Navier-Stokes-Fourier hydrodynamics predict instabilities. It may
be that this is not necessarily the case, since the linearized
version of the transport equations used in those works introduce
the heat flow in the stress-energy tensor. The well known
Rayleigh-Brillouin (RB) spectrum, corresponding to damped
perturbations can be  calculated from the linearized equations and
verified by experiment. It will be part of future work the
computation of the relativistic counterpart of the RB spectrum
with the version here presented of relativistic irreversible
thermodynamics. The modifications by gravity to the RB spectrum in
the Newtonian case has already been investigated \cite{Nos2}
\cite{Nos3}.  It is also interesting to consider other
relativistic hydrodynamical systems such as the quark-gluon plasma
\cite{Elze}.

In our opinion much more work is needed to fully access the
physical content of the relativistic Meixner theory before it is
virtually discarded incorporating arguments of other thermodynamic
approaches whose physical content, even in its non-relativistic
version, is still rather controversial. More evidence along these
lines will be given in the next future.

 This work has been partially supported by CONACyT (Mexico),
project 41081-F.


\begin{thebibliography}{10}


\bibitem[1]{Meixner1} J. Meixner; Ann. Physics \textbf{41} 409
(1942); ibid \textbf{43}, 244 (1943) and Z. Phys. Chem.
\textbf{43}, 244 (1943).

\bibitem[2]{Groot1} S.R. de Groot, Thermodynamics of irreversible
processes (North Holand publishing Co. Amsterdam 1952) Chap. XI is
a very valuable reference for these purposes (1970).

\bibitem[3]{kin1} S.R. de Groot and P. Mazur, Non-equilibrium
thermodynamics (reprinted by Dover publications Inc. Mineola N.Y.
1984).

\bibitem[4]{Steele} W.A. Steele; Time correlation functions in
transport phenomena in fluids; H.J.M. Hanley, editor (M. Dekker,
New York 1969) Chap. 8

\bibitem[5]{kin2} J.O. Hirschfelder, C.F. Curtiss and R.B.
Bird, The molecular theory of liquids and gases (John Wiley ans
Sons, N.Y. 2nd Ed. 1964).

\bibitem[6]{GC1} P. Golstein and L.S. Garcia-Colin,
"On the validity of Onsager's relations in an inert dilute gas
mixture", J.Non-equilib. Thermodyn. 2005 (in press).

\bibitem[7]{GC2} For  a review see: L.S. Garcia-Colin and
F.J. Uribe, Extended irreversible thermodynamics beyond the linear
regime: a critical overview. J.Non-equilib. Thermodyn.
\textbf{16}, 89 (1991).

\bibitem[8]{Jou1} D. Jou, G. Lebon and J. Casas-Vazquez,
Extended irreversible thermodynamics (Springer Verlag, N.Y. 2nd.
Ed. 1996).

\bibitem[9]{GC3} L.S. Garcia-Colin, Extended irreversible
thermodynamics, an unfinished task. Mol. Phys. \textbf{86} 997
(1995).

\bibitem[10]{Jou2} D. Jou, J. Casas-Vazquez and G. Lebon,
Rep. Prog. Phys. \textbf{62} 1055 (1999) and references cited
therein.

\bibitem[11]{Nos1} A. Sandoval-Villalbazo and L.S. Garcia-Colin,
Relativistic Navier-Stokes equations in the Meixner-Prigogine
scheme. Physica A \textbf{240} 480 (1997).

\bibitem[12]{Eckart} C. Eckart, The thermodynamics of irreversible
processes I: the simple fluid. Phys. Rev. \textbf{58}, 267 (1940);
II: fluid mixtures; ibid \textbf{58},919 (1940). III: The
relativistic theory of the simple fluid; ibid \textbf{58}, 919
(1940).

\bibitem[13]{Tolman} R.C. Tolman, Relativity, thermodynamics and cosmology
(Oxford University Press, Oxford  1934) Chap IX

\bibitem[14]{Weyl} H. Weyl: Space Time and Matter, Dover
Publications Inc. NY 1950) Chap. IV.

\bibitem[15]{Israel1} W. Israel, Nonstationary Irreversible Thermodynamics:
a Causal Relativistic Theory, Ann. Phys (N.Y.) \textbf{100} 310
(1976) and references therein.

\bibitem[16]{Israel2} W. Israel, Thermodynamics of relativistic
systems, Physica \textbf{106A} 209 (1981).

\bibitem[17]{Israel3} W. Israel and J.M. Stewart, progress in
relativistic thermodynamics and electrodynamics of continuous
media in "General relativity and gravitation" Vol 2; A. Held,
editor (Plenum press, N.Y. 1980).

\bibitem[18]{Zimdahl} W. Zimdahl, Bulk Viscous Cosmology; Phys.
Rev. D \textbf{53}, 5483 (1996).

\bibitem[19]{Jou4} D. Pavon, D. Jou and J. Casas-Vazquez, Ann.
Inst. Henri Poincare \textbf{A36} 79 (1982).

\bibitem[20]{Hiscock} W.A. Hiscock and L. Lindblom; Generic
instabilities in first order dissipative relativistic fluids,
Phys. Rev. D \textbf{31} 725 (1985).

\bibitem[21]{GC4} F.J. Uribe, R.M Velasco and L.S. Garcia-Colin;
Burnett description of strong shock waves, Phys. Rev. Lett.
\textbf{81}, 2044 (1998).

\bibitem[22]{GC5} F.J. Uribe, R.M Velasco, L.S. Garcia-Colin and E. Herrera;
Shock Waves Profiles in the Burnett Approximation, Phys. Rev. E.
\textbf{62}(2000).

\bibitem[23]{Israel4} W. Israel, Relativistic thermodynamics,thermofield statistics and
superfluids, J. Non-equlib. Thermodyn. \textbf{295} (1986) (see
section 4).


\bibitem[24]{Nos4} A. Sandoval-Villalbazo and L.S. Garcia-Colin,
On the cosmological implications of irreversible thermodynamics,
J. General Relat. and Grav. \textbf{31}, 781 (1999).


\bibitem[25]{Nos2} A. Sandoval-Villalbazo and L.S. Garcia-Colin,
Hydrodynamic time correlation functions in the presence of a
gravitational field, Physica \textbf{A327}, 213  (2003).

\bibitem[26]{Nos3} A. Sandoval-Villalbazo and L.S. Garcia-Colin,
Jeans instability in the linearized Burnett regime, Physica
\textbf{A347}, 375 (2005).

\bibitem[27]{Elze} H.-Th. Elze, J. Rafelski and L. Turko, Entropy
Production in Relativistic Hydrodynamics, Phys.Lett. \textbf{B506}
123 (2001).

\end{thebibliography}
\end{document}